# Adatom pair distribution up to half coverage: O-Pd(100)

Wolfgang Kappus

Alumnus of ITP
Philosophenweg 19
D-69120 Heidelberg

wolfgang.kappus@t-online.de

## Abstract

The superposition approximation and the Born-Green-Yvon (BGY) equation allows calculation of pair distributions in thermal equilibrium from pair potentials. A 2-d variant can be used to derive adatom pair distributions from arbitrary analytical pair potentials. As practical example substrate mediated elastic interactions, fitted previously to first principles (FP) calculations, are used to derive adatom pair distributions of O-Pd(100). The evaluation method utilizes the particle-hole symmetry of the pair interaction lattice gas Hamiltonian. The nonlinear BGY-type integral equation is solved numerically up to half coverage using power series expansion of coverage.

The resulting adatom pair- and three-body distributions are used to analyze the short-range order of adatoms. A Monte Carlo (MC) simulation was performed in parallel to compare the current model with established methods. MC derived pair distributions fit well to those found with the analytical model, except in the high coverage region. The Fourier transformed pair correlations are compared with experimental LEED spot data and provide additional insight.

Due to significant attraction the model predicts formation of 3rd and 4th nearest neighbors (n.n.) at small 3rd n.n. chains. A long range order of adatoms like a p(2x2) lattice is not found with the current model but a glassy structure is proposed as also indicated by the equilibrium Monte Carlo configurations. The glassy structure is assumed to contain fluctuations explaining broad p(2x2) LEED spots in the 0.25 monolayer region. The assumptions and limitations of the model towards higher coverages are discussed and open questions are formulated.

## Keywords

# 1. Introduction

Interactions of adatoms are subject of continuous interest, various different interaction mechanisms have been described in detail [1]. Adatom structures like superlattices, nanodot arrays, nanostripes, strain relief patterns are interesting for various general and technological reasons; reviews were given in [2,3,4,5]. Lateral interactions also govern the ordering behavior of adatoms and thus influence the catalytic activities of surfaces [6]. Lattice gas models are the discrete representations of surface properties and usable for predicting growth and structure of adatom configurations like adatom diffusion and phase diagrams [1].

Monte Carlo (MC) simulations are established to compute equilibrium or dynamic properties of e.g. adatom configurations from their lateral interactions [7]. It will be shown in the following sections that the direct calculation of adatom pair distributions can provide additional insight to the order structure of adatoms.

The adatom pair distribution analysis, together with its related three-body distribution through the superposition approximation, will allows one to derive the adatom neighborhood at various coverages. It is based on the well known superposition approximation allowing to formulate the Born-Green-Yvon (BGY) equation [8], adapted to the 2-dimensional case [12]. The equation is also usable for a spatially discrete system of adsorption sites. This nonlinear integral equation links the adatom pair potential to the adatom pair distribution as function of the coverage. To utilize this equation, the pair potential needs to be analytic because its derivative is part of the equation. Any analytical pair potential is usable, in the most simple case an isotropic power law. Adatom interactions with both repulsive and attractive parts can be described e.g. by an oscillating potential.

The O-Pd(100) system was selected as an example because it has repulsive and attractive interactions within its 4 nearest Oxygen neighbors, because an analytical potential was available [10] and because O-Pd(100) has been studied extensively. Furthermore O-Pd is technically used e.g. in cars as oxidizing catalyst. Surfaces of metals like Palladium are a topic of ongoing research, see [11] and references therein.

Thanks to computational power and advanced algorithms the free energies of adatom configurations have been determined from first principles using density-functional theory (DFT) and used as inputs to derive lattice-gas Hamiltonians (LGHs) with cluster expansion (CE) techniques [9]. This method was applied to the O-Pd(100) system to model surface ordering of atomic adsorbates on Palladium [13].

Instead of using CE technique, leading to many pair-, trio-, and quarto interaction parameters, the DFT results of [13] for the O-Pd(100) system were differently interpreted in terms of elastic adatom interactions and an appropriate lattice gas Hamiltonian was derived using just 3 parameters for pair- and many-body interactions [10]. Key feature of this elastic Hamiltonian is its analytic nature allowing differentiation and thus is a candidate for a pair distribution analysis based on the 2d Born-Green-Yvon (BGY) type equation [12]. This method complements Monte Carlo simulations for evaluating the system partition function and order parameters like in [9].

While the modern ab-initio calculations concentrate on short range interactions of adatoms, the mesoscale range is a domain of superimposed fields since decades. The classic $s^{-3}$ elastic interactions e.g. are caused by substrate strain fields, the derivatives of displacement fields. The pair distribution field, derived from the pair interaction field utilizing the 2d BGY equation, is just another mesoscale example.

In the subsequent pair distribution analysis up to 0.5 monolayer (ML), many-body interactions are ignored because they are marginal in the O-Pd(100) case [10, 11] and because surface reconstruction starts at 0.4 ML coverage [14]. Avoiding many-body interactions allows making use of the particle-hole symmetry of the lattice gas Hamiltonian [15]. The BGY-type equation has to be adapted to this symmetry, also extending the reach of BGY which ignores 4-party and higher distributions.

The adatom pair distribution analysis, together with its related three-body distribution through the superposition approximation, will allow to derive the adatom neighborhood at various coverages and temperatures. MC simulated adatom configurations at certain coverages will be helpful to illustrate the short-range order of adatoms. The current BGY based analysis does not cover multi-body distributions beyond 3 adatoms; fluctuations with additional order are to be expected.

The O-Pd(100) system was studied intensively experimentally, see [16] and references therein. LEED measurements formed the basis for establishing a model for surface phases. The Fourier transform of the adatom correlation will allow comparisons with LEED pattern measurements of O-Pd(100) [16] and in consequence will put some earlier conclusions about surface phases in question [16, 17]. A glassy structure imbedding ordered microstructures is proposed under the condition of a significant 4th nearest neighbor attraction.

The paper is organized as follows:
After outlining the general motivation in section 1, the pair distribution analysis method is detailed in section 2. The elastic interaction model is recalled in section 3. In section 4 the pair distribution analysis results are presented and adatom ordering is outlined. Monte Carlo simulation results are added for comparison. In section 5 applicability and limitations of the model are discussed and open questions are addressed. Section 6 closes with a summary of the results.

## 2. Short-range analysis

A direct statistical method will be used in this section to analyze the short-range order of adatoms for varying coverages and temperatures. The adatom-vacancy symmetry for pair interactions will be utilized to propose an extension to a known BGY-type equation relating adatom interaction, coverage and pair distribution. Results using the Monte Carlo method will be shown for comparison. The frequently used Monte Carlo method determines thermodynamic functions by averaging over many simulation passes.

2.1. Adatom pair distribution calculation method

The probability for finding an adatom at a certain lattice site equals the coverage $\theta$. If adatoms are correlated e.g. due to an interaction, the probability $pr$ to find an adatom at position $s_2$ if another is located at position $s_1$ is the product of coverage and the pair distribution $g$

$$pr(s_1, s_2) = \theta g(s_1, s_2). \quad (2.1)$$

When the three-body configurational distribution for atoms interacting in a volume is expressed in terms of the pair distribution, Kirkwoods superposition approximation [18]

$$g(s_1, s_2, s_3) = g(s_1, s_2) g(s_2, s_3) g(s_3, s_1), \quad (2.2)$$

then the pair distribution $g_{12} = g(s_1, s_2)$ in thermal equilibrium is given by the Born-Green-Yvon integro-differential equation [8], introduced to describe dense fluids with the pair interaction of atoms $u_{12} = u(s_1, s_2)$ and its gradient $\nabla u_{12}$

$$\nabla^{(1)} [\ln g_{12} + u_{12}] = -\rho \int_V g_{23} g_{13} \nabla^{(1)} u_{13} \, d\mathbf{r}_3 . \quad (2.3)$$

Here $u_{12} = U_{12}/k_B T$ denotes the scaled pair interaction, $\rho$ is the density, $\nabla^{(1)}$ acts on the coordinates of atom (1) only. The left hand side of (2.3) describes the familiar zero coverage Boltzmann distribution while the right hand side of (2.3) introduces the nonlinear density dependent correlation.

Back to the 2-d case Eq. (2.3) can be transformed into an integral equation for $g_{12}$ [12]

$$\ln g_{12} + u_{12} = \frac{1}{2\pi} \theta \int_S (g_{23} - 1) \, d\mathbf{s}_3 \int_S g_{43} \frac{s_{14} \nabla^{(4)}}{s_{14} s_{14}} u_{43} \, d\mathbf{s}_4 \quad (2.4)$$

with $\theta$ denoting the adatom coverage. Eq. (2.4) proved solvable numerically for low coverages at medium temperatures. To extend the reach of the pair distribution analysis towards half coverage we utilize the adatom-vacancy symmetry for adatom pair interactions at half coverage [15]

$$\bar{g}_{12}(0.5) = g_{12}(0.5) \qquad (2.5)$$

with $\bar{g}$ denoting the vacancy pair distribution function, in the pair interaction case following

$$\ln \bar{g}_{12} + \bar{u}_{12} = \frac{1}{2\pi} \bar{\theta} \int_S (\bar{g}_{23} - 1)\, d\,s_3 \int_S \bar{g}_{43} \frac{s_{14}\, \boldsymbol{\nabla}^{(4)}}{s_{14}\, s_{14}} \bar{u}_{43}\, d\,s_4, \qquad (2.6)$$

where

$$\bar{\theta} = 1 - \theta, \quad \bar{u}_{12} = -u_{12}, \quad \boldsymbol{\nabla}^{(4)} \bar{u}_{43} = -\boldsymbol{\nabla}^{(4)} u_{43}. \qquad (2.7)$$

A new Eq. (2.8) can be found fulfilling Eq. (2.5) and interpolating Eqs. (2.4) and (2.6) smoothly, simultaneously preserving the $\theta \to 0$ limit in Eq. (2.4) and the $\theta \to 1$ limit in Eq. (2.6):

$$\ln g_{12} = -\cos^n(\pi\theta)\, u_{12} + \frac{1}{2\pi} \frac{1}{\pi} \sin(\pi\theta) \int_S (g_{23} - 1)\, d\,s_3 \int_S g_{43} \frac{s_{14}\, \boldsymbol{\nabla}^{(4)}}{s_{14}\, s_{14}} u_{43}\, d\,s_4\,. \qquad (2.8)$$

Eq. (2.8) simultaneously approximates the adatom pair distribution $g_{12}$ up to half coverage and the vacancy pair distribution in the range beyond. It must be noted that this interpolation is not unique, the exponent $n$ of the $\cos^n$ term could be chosen e.g. 1 or 3, other functions could be taken as well. The solution $g(\theta,s)$ is Eigenfunction of Eq. (2.8).

2.2. Power series expansion

The integral equation (2.8) for $g_{12}$ is a numerical challenge because it is extremely nonlinear. It can be solved using a power series expansion technique. Starting from zero, coverage is incremented by small steps of e.g. $d\theta=0.005$ until half coverage. Each calculated pair distribution is input for the subsequent step.

The zero coverage distribution is

$$g_{12}(0) = \exp(-u_{12}). \qquad (2.9)$$

Applying a power series expansion in $\theta$, the pair distribution (omitting indices for adatoms 1 and 2) $g(\theta + d\theta)$ is

$$g(\theta + d\theta) = g(\theta) + d\theta\, g' + \frac{1}{2} (d\theta)^2\, g'' + \ldots \qquad (2.10)$$

We denote the right hand side of Eq. (2.8) as $F(\theta,g(\theta))$, so

$$g(\theta) = \exp(F(\theta, g(\theta)))\,, \qquad (2.11)$$

which gives

$$g'(\theta) = g(\theta)\, F'(\theta, g(\theta))\,, \quad \text{etc.} \qquad (2.12)$$

Differentiation of $F$ is straightforward. In the current calculation the series Eq. (2.10) was used up to the $(d\theta)^2$ term.

2.3. Superimposed fields in a lattice gas environment

The application of a BGY-type equation to a lattice gas model needs explanation, because the BGY equation was used for the description of fluids in statistical physics. The BGY equation defines the pair distribution field from the interaction field and hereby makes use of an interaction field derivative. Construction of an interaction field from a few discrete adatom interactions on a lattice gas (calculated by first principles) would mean ambiguous extrapolation with doubtful meaning of derivatives.

If, however, an interaction field is derived from elasticity theory it has a solid base. It is mediated by substrate strain fields, the derivative of displacement fields. Consequently also derivatives of the interaction field are meaningful. So the introduction of pair distributions as further field for the description of adatom configurations appears valid.

In [10] it was shown that adatom interactions based on elasticity theory with a few parameters fit well to adatom configuration energies of Oxygen on Pd(100) calculated by first principles [13]. The interaction found in [10] is comparable to the classical $s^{-3}$ elastic interaction but shows oscillations. The interaction values for discrete separations are unique and are usable for MC simulations and for other stochastic methods. The derivatives in the BGY formulation represent forces between adatoms at discrete sites (balanced in equilibrium). Using this potential allows to include interactions beyond 5th NN in lattice sums, keeping the interactions of narrow positions comparable to other lattice gas approaches.

### 2.4. Lattice algorithm

Eqs. (2.8), (2.9), (2.10), (2.11), (2.12) can be solved numerically for arbitrary analytical pair interactions $u_{12}(s_1,s_2)$ if the continuous functions $u_{12}, g_{12}, s_{12}$ are replaced by their values at surface lattice points with the grid coordinates $(i,j)$ of adatom 1 and coordinates $(k,l)$ of adatom 2. The pair distribution of those two adatoms then reads $g_{ij,kl}$ and the area integrals in Eq. (2.8) have to be replaced by proper sums

$$\ln g_{ij,00} = -\cos^n(\pi\theta)\, u_{ij,00} + \frac{1}{2\pi^2}\sin(\pi\theta)\sum_{k,l}(g_{00,kl}-1)\sum_{m,n} g_{mn,kl}\frac{\mathbf{s}_{ij,mn}\,\mathbf{\nabla}^{(mn)}}{s_{ij,mn}\,s_{ij,mn}}u_{mn,kl}\ . \quad (2.13)$$

In short notation $g_{ij,00}$ can be written as $g_{ij}$. The pair distributions $g_{ij,00}(\theta)$ between adatoms obeying Eq. (2.13) are Eigen matrices. The algorithm to calculate them from the zero coverage matrix $g_{ij}(0)$ is given in section 2.2.

### 2.5. Fourier Transform

To enable a comparison of calculated pair distributions with LEED spots of the O-Pd(100) system, the Fourier Transform $\tilde{v}$ of the correlation $v_{12}=g_{12}-1$ will be calculated. The correlation $v_{12}(s)$ between two adatoms is expected to decrease rapidly as their distance $s$ increases, while $g_{12}(s)$ approaches 1 for large distances. The Fourier Transform $\tilde{v}$ at certain points of the reciprocal lattice correspond to LEED spots and thus allows comparison with experiments.

### 2.6. Three body distribution

The three-body configurational distribution in the superposition approximation

$$g^{(3)}_{ik,mn} = g_{ik}\, g_{mn}\, g_{m-i,n-k} \qquad (2.14)$$

relates the three adatoms located at the grid coordinates $(i,k),(0,0),(m,n)$. It will allow to inspect adatom configurations complementing two-body distributions and Fourier Transforms. The three-body distribution derived using Eq. (2.14) is a tensor $\mathbf{g}^{(3)}=[g_{ik,mn}]$.

### 2.7. Monte Carlo simulation

A Metropolis type Monte Carlo simulation was performed to derive adatom pair distributions from thermal equilibrium configurations. The lattice size was 30 x 30, periodic boundary conditions were applied. Unlike in the analytical method above, the MC method requires the potential to be truncated. Nevertheless this will allow to compare the the analytical distributions with statistical distributions.

# 3. Adatom Interaction Model

An eigenvector model describing substrate mediated elastic interactions on Pd(100) between oxygen adatoms [10] is used in the following as example for the following reasons:
- it is analytic and using its derivatives allows direct calculation of Eq. (2.13)
- it is usable for describing a lattice-gas
- it interprets First Principles calculations of the free energy of O-Pd(100) configurations [13] with just 3 free

parameters [10]
- it comprises pair interactions; many-body interactions necessary to fit the First Principles results are marginal and not relevant up to half coverage
- it reflects the DFT verifications [19] that O atoms are 4-fold coordinated, that the Pd(100) substrate is distorted when O atoms are adsorbed and that the O-Pd bonds are electronically localized, i.e. that the assumptions of the present elastic model are valid
- it describes adatom-adatom interactions both repulsive and attractive within the 4 nearest neighbors.

In the subsequent section the results of [10] are recalled, restricted to pair interactions of adatoms because
- many-body interactions are marginal
- the pair distribution analysis of section 2 covers only pair interactions
- strong nearest neighbor repulsion of oxygen atoms on Pd(100) hinders the formation of nearest neighbor pairs below half coverage. Such pairs are responsible for many body effects at higher coverage
- O-Pd(100) shows surface reconstruction beyond half coverage [14] not covered by the current flat surface model.

The general interaction $U_{kl}(s,\chi)$ of [10] therefore is restricted to the pair interaction $U_{11}(s,\chi)$, for simplicity denoted $U(s,\chi)$ within the current analysis.

### 3.1. Elastic interaction model

The elastic energy of a substrate with adatoms in a continuous description is given by the sum of two parts, the energy of the distorted substrate and the energy of adatoms exerting tangential forces on the substrate

$$H_{el} = \frac{1}{2} \int_V \epsilon(r)\, c\, \epsilon(r)\, dr + \int_S \epsilon(s)\, \pi(s)\, ds. \qquad (3.1)$$

Here $\epsilon = [\epsilon_{\alpha\beta}]$ denotes the strain tensor field, $c = [c_{\alpha\beta\mu\nu}]$ denotes the elastic constants tensor, and $\pi = [\pi_{\mu\nu}]$ denotes the force dipole- or surface stress tensor field. The integrals comprise the bulk V or the surface S.

Following [10], using a diagonalization method and replacing the continuous fields by a lattice description, the elastic energy of $m$ adatoms can be written

$$H_{el} = \sum_{i=1}^{m} \sum_{j=2}^{i} U(s_i, s_j)\, n_i n_j \qquad (3.2)$$

with $n_i \in \{0, 1\}$ denoting occupation numbers of lattice positions $s_i$ (in $s_0$ units) and the elastic pair interaction $U(s_i, s_j)$ approximated by

$$U(s, \chi) = (2\pi)^{-1} \frac{P^2}{c_{44}} \sum_p \hat{\omega}_p \cos(p\chi) \cos(p\pi/2)\, 2^{-1-p} \kappa_B{}^3 (s\kappa_B)^p\, \Gamma\!\left(\frac{3+p}{2}\right) *$$

$$_pF_q\!\left(\left(\frac{3+p}{2}\right), \left(\frac{5+p}{2}, 1+p\right), -\frac{1}{4} s^2 \kappa_{BZ}{}^2\right) \Big/ \left(1 + (s/s_0)^{3/2}\right), \qquad (3.3)$$

where $\Gamma(p)$ denotes the Gamma function and $_pF_q(a,b,z)$ the generalized hypergeometric function; $p$ attains the values $\{0, 4, 8\}$, $\chi$ is the angle between the $x$- and the $s_j - s_i$ direction; $\kappa_{BZ}$ is the surface Brillouin zone cutoff parameter of the order $2\pi/s_0$; the $\hat{\omega}_p$ are dimensionless constants. $U(s,\chi)$ is oscillating and decays with $s^{-3}$. The pair interaction is proportional to the square of the surface stress parameter $P$ and inverse proportional to the elastic constant $c_{44}$.

Since Eq. (3.3) looks complicated a simpler example is recalled:
In the case of an elastic isotropic substrate (e.g. tungsten) and an isotropic stress tensor field only the $\hat{\omega}_0$ term does not vanish

$$\hat{\omega}_0 \text{ (isotropic)} = -c_{11} / (2(c_{11} - c_{44})) \quad . \qquad (3.4)$$

and in the long range limit of the isotropic pair interaction follows a repulsive $s^{-3}$ law [20]

$$U(s) \to -(2\pi)^{-1} \frac{P^2}{c_{44}} \hat{\omega}_0 \, s^{-3} . \qquad (3.5)$$

Also $U(s,\chi)$ in Eq. (3.3) shows a $s^{-3}$ long range (oscillating) tail. The $\hat{\omega}_p$ values of the anisotropic Pd(100) (using elastic constants $c_{11}$=221, $c_{12}$=171, $c_{44}$=70.8 GPa [21]) shows Tab. 1.

| $\hat{\omega}_0$ | $\hat{\omega}_4$ | $\hat{\omega}_8$ |
|---|---|---|
| −0.9431 | 0.1009 | 0.0016 |

Table 1. Coefficients $\hat{\omega}_p$ on Pd(100).

The parameters $\kappa_B$=5.85/$s_0$ and $P$=9.505 meV in Eq.(3.3) for O-Pd(100) are taken from [10], used for fitting the FP results of [13]. It must be noted that the numerical value for $P$ was derived under the convenience assumption $c_{44}s_0^3$=1 meV for allowing a direct evaluation of Eq. (3.3).

$U(s,\chi)$ as defined in Eq. (3.3) and with the above parameters for O-Pd(100) shows repulsion for the nearest (228 meV) and 2nd nearest neighbors (66 meV) and small attraction for the 3rd (-43 meV) and 4th nearest neighbors (-15 meV) [10]. Those values can be compared with the TPD spectra results of [11] with nearest (368 meV), 2nd nearest (130 meV), 3rd nearest (-40 meV), 4th nearest neighbor (-2 meV). The difference can be explained by the long range of $U(s,\chi)$ in contrast to just 4 nearest neighbors in [11]. The $U(s,\chi)$ tail significantly contributes to the lattice sum and thus lowers the nearest neighbor interactions.

The size of the 4th nearest neighbor interaction has significant impact on adatom order. A lattice model with 4 nearest neighbors and the 4th nearest neighbor attraction of only -2 meV would generate a strong p(2x2) phase near 0.25 ML coverage [11].

# 4. Pair distribution results and short-range order analysis for O-Pd(100)

In this section we make use of Eq. (3.3), describing the analytical adatom-adatom interaction $U(s, s')$. Oxygen adatom positions $s_i$ are assumed on fourfold coordinated position of the Pd(100) substrate, forming a grid. For convenience we will use the notations p(2x2) and c(2x2) for ordered phases to also name low ordered adatom configurations. p(2x2) stands for a 0.25 ML phase, c(2x2) for the half coverage phase. The $g_{ik}$ notation for the pair distribution follows the one introduced in section 2.4.

### 4.1. Adatom pair distribution results

The pair distribution according to Eq. (2.1) is a measure for the probability to find an adatom at position $s_2$ if another is located at position $s_1$. Evaluating Eqs. (2.13) and (2.9,10,11,12) with the pair interaction of Eq. (3.3) results in a $g_{ik}$ matrix for lattice positions $(i,k)$ of adatom 2, if adatom 1 is located at $(0,0)$. While further distant adatoms approach $g_{ik}$=1, the near neighbors values differ from 1.

Fig. 1 shows for the temperature 400 K $g_{ik}(\theta)$ for the nearest to the 5th nearest neighbors obeying Eq. (2.13) with the exponent $n$=3. The slope of the attractive 3rd nearest neighbors reaches a maximum near 0.1 ML coverage and then continuously decreases. The maximum height of nearly 4 indicates considerable ordering of adatoms. The attractive 4th nearest neighbors slope continuously decreases. The other slopes represent repulsive neighbors. The 2nd nearest neighbor slope slowly increases towards 1. The nearest neighbor slope begins to rise from almost zero at 0.2 ML coverages and approaches 1.26 at half coverage. The 5th nearest neighbor first decreases slightly and rises again from 0.2 ML coverage. We note that $g_{ik}(0.5)$ differs from 1, the trivial solution of Eq. (2.13).

According to Eq. (2.1) $\theta * g_{ik}(\theta)$ is the probability of finding an adatom at position *(i,k)* if another is located at position *(0,0)*. The probability of finding e.g. a 3rd nearest neighbor at 0.1 ML coverage thus is about 0.4 in this calculation.

Fig. 1 shows for low coverages a considerable amount of adatom pairs at 400 K. At 0.1 ML coverage the probability for finding isolated adatoms is 0.1; for ((0,0),(2,0)) 3rd neighbor pairs its about 0.4; for ((0,0),(2,1)) 4th nearest neighbor pairs its about 0.14. Pairs may, of course, be parts of clusters. With increasing coverage and building of larger clusters the picture becomes more complicated - the three body distributions below will help to clarify.

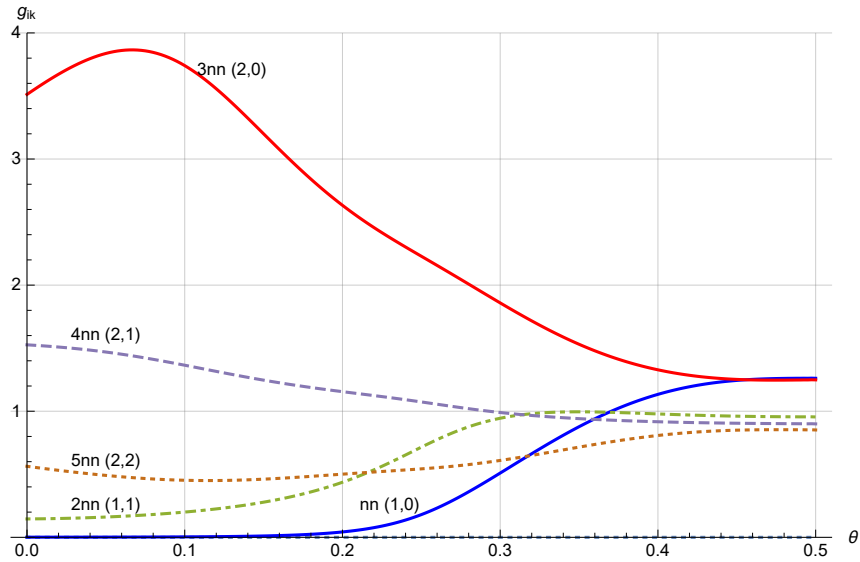

Fig. 1: Adatom pair distribution $g_{ik}(\theta)$ at 400 K from nearest to 5th nearest neighbors *(i,k)* of *(0,0)* in dependence of coverage $\theta$ obeying Eq. (2.13).
Red line: 3rd n.n. (2,0), broken line 4th n.n. (2,1), dotted line: 5th n.n. (2,2), dot broken line: 2nd n.n. (1,1), blue line: n.n. (1,0).

### 4.2. Monte Carlo simulation results

Simulations of O-Pd(100) with 10 nearest neighbor interactions have been performed at 10 different coverages. Due to limited computing resources the variances of $g_{ik}(\theta)$ were up to 4% at half coverage, up to 20% at 0.1 ML and up to 100% for the small $g_{ik}(\theta)$ values at 0.05 ML.

Fig.2 shows for the temperature 400 K $g_{ik}(\theta)$ for the nearest to the 5th nearest neighbors calculated using the Monte Carlo method. Coverages in 0.05 ML steps have been evaluated. Comparison with Fig.1 shows almost the same pair distribution. Main difference is the weaker increase of the n.n. curves at high coverage. The limits of the analytical model and of the interpolation scheme of Eq. (2.13) becomes obvious at high coverages. We also recall the shorter range of the interaction used for the MC simulation.

The additional value of MC calculations is visible by adatom configurations at various coverages:
- at 0.1 ML many 3rd and 4th nearest neighbor pairs are found, sometimes forming irregularly formed clusters
- but no grid; the length of 3rd n.n. chains does not exceed 3 adatoms
- at 0.25 ML (the value for p(2x2) grids) a statistical mixture of 3rd and 4th nearest neighbors dominates, some 2nd n.n. adatoms occur as well, a few 3rd n.n. chains with 3 to 4 adatoms are also found
- at 0.5 ML (the value for c(2x2) grids) the lines of 2nd n.n. adatoms (or vacancies) are irregularly broken by n.n. links. The length of 2nd n.n. chains seldom exceeds 4 adatoms.
All configurations thus provide helpful visualizations of a glassy adatom structure.

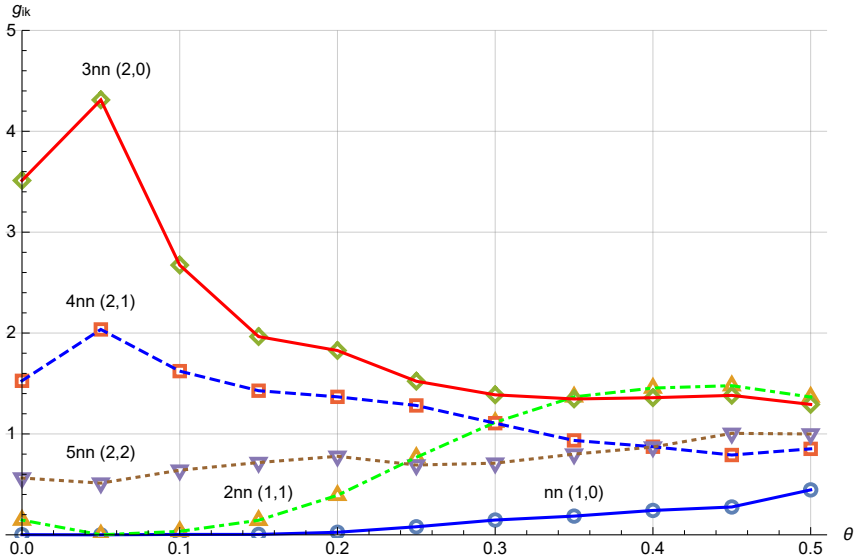

Fig. 2: Monte Carlo simulated adatom pair distribution $g_{ik}(\theta)$ at 400 K from nearest to 5th nearest neighbors *(i,k)* of *(0,0)* in dependence of coverage $\theta$ calculated in 0.05 ML steps. $g_{ik}(\theta)$ values marked by different symbols, connected by lines for guidance. Red line: 3rd n.n. (2,0), broken line 4th n.n. (2,1), dotted line: 5th n.n. (2,2), dot broken line: 2nd n.n. (1,1), blue line: n.n. (1,0).

4.3. Fourier analysis results and comparison with LEED measurements

LEED spots of adatom grids with long range order are sharp; with decreasing order they get broader. On O-Pd(100) two spots (relevant for this analysis) have been reported for coverages up to 0.5 ML, p(2x2) indicating a grid of 3rd nearest neighbors at 0.25 ML coverage and c(2x2) indicating a grid of 2nd nearest neighbors at 0.5 ML coverage [16].

A simple model for O-Pd(100) would consist of p(2x2) domains coalescing towards 0.25 ML coverage and filling 2nd nearest neighbor positions beyond 0.25 ML until the c(2x2) grid has formed at half coverage. Reality is less simple [16]. The Fourier transform of the correlation is in a 3-d plot a landscape with planes and mountains, latter representing LEED spots.

Fig. 3. shows for the temperature 400 K the corresponding Fourier transformed correlation maxima $\tilde{\nu}(\theta)=\text{FT}(g(\theta)-1)$ at two meaningful points of the reciprocal lattice (units are arbitrary) :
 - $\tilde{\nu}(1/2,1/2)$, representing both c(2x2) and p(2x2) spots, red straight line
 - $\tilde{\nu}(0,1/2)$, representing a p(2x2) spot, yellow dashed line.
The difference is shown as a dotted line.
The spot maxima (proportional to their volume) provide measures for LEED spot intensities.
Both lines show a maximum near 0.1 ML coverage indicating some kind of ordering; then they decrease until 0.3 to 0.35 ML coverage and rise again slightly at higher coverage, again indicating some ordering. The oscillations near half coverage indicate the onset of numerical instability.

It must be noted that the (1/2,1/2) spot at low coverages is broad and skew, containing 4th nearest neighbor patterns. The dotted difference line therefore does not represent a c(2x2) grid.

Comparing Fig.3 with the LEED spot measurements at 400 K in Fig. 4 of [16] we conclude a similar slope for the (1/2,1/2) spot shifted to lower coverages. The (0,1/2) spot intensity looks comparable to this intensity in [16]. The authors of [16] suppose a single (2x2) phase with a gradually increasing number of adatoms at 2nd nearest neighbor sites. The experimental setup with indirect coverage calibration and bulk dissolution implies variances. Non equilibrium effects must also be considered.

Fig. 3. shows a slight increase of both the (1/2,1/2) and the (0,1/2) slopes beyond 0.4 ML coverage, not visible in [16]. Such increase is also not visible if the interpolation scheme of Eq. (2.13) is changed towards smaller

values of the coverage dependent term. This shows the limits of the model in the high coverage range.

Summarizing, the broad Fourier transform/ LEED spot intensities do not show a long range order of adatoms at 400 K. The LEED short-range order measurements at similar temperatures in [16] have been interpreted by longer range ordered phases. This interpretation is put in question with the current model where long range interactions and a significant 4th n.n. attraction are present.

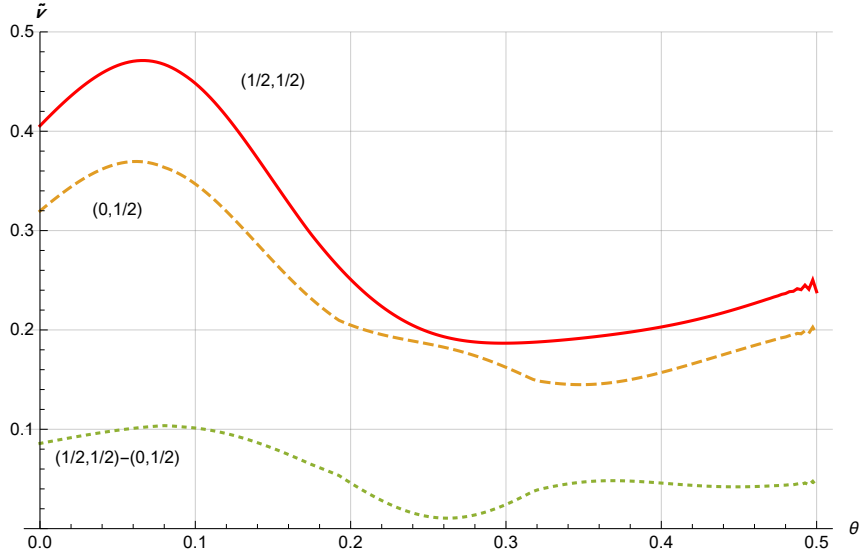

Fig. 3: Fourier transformed correlation $\tilde{v}(\theta)$ (LEED spot intensities) at 400 K for representative points of the reciprocal lattice $(j,l)$ in dependence of coverage $\theta$. Pair distribution $g_{ik}(\theta)$ obeying Eq. (2.13). Red line: (1/2,1/2), brown broken line (0,1/2), dotted line shows their difference.

4.4. Adatom 3-body distribution

The Kirkwood approximation of Eq. (2.2) allows deriving 3-adatom distributions $g_{123}$ providing additional insight on equilibrium cluster formation especially in the lower coverage region. Fig. 4 shows $g^{(3)}_{ik,mn}(\theta)$ of representative trios $(i,k),(0,0),(m,n)$ for the temperature 400 K evaluated with Eq. (2.14).

At all coverages 3rd n.n. straight $(2,0)(0,0)(\bar{2},0)$ trios dominate, followed by 3rd n.n./4th n.n. $(2,0)(0,0)(1,2)$ 63° trios. 3rd n.n. $(2,0)(0,0),(0,2)$ 90° trios and 3rd n.n./4th n.n. $(2,0)(0,0)(\bar{1},2)$ 116° trios are closely below.

$(1,0)(0,0)(\bar{1},0)$ n.n. trios show up beyond 0.3 ML coverage. They indicate n.n. faults in a perfect c(2x2) grid. So near half coverage the formation of a longer range c(2x2) grid shows unlikely.

Near 0.1 ML coverage the value of $g^{(3)}_{ik,mn}(\theta)$=12 for the 3rd n.n. straight $(2,0)(0,0)(\bar{2},0)$ trios mean 12 times the uncorrelated value of 1, significantly higher than the values of other trios around 7 to 8. This indicates adatom formation preferably at short 3rd n.n. chains as stated in the MC section above.

The high amount of 3rd n.n./4th n.n. 116° and 63° trios at 0.25 ML makes the formation of a longer range p(2x2) grid unlikely.

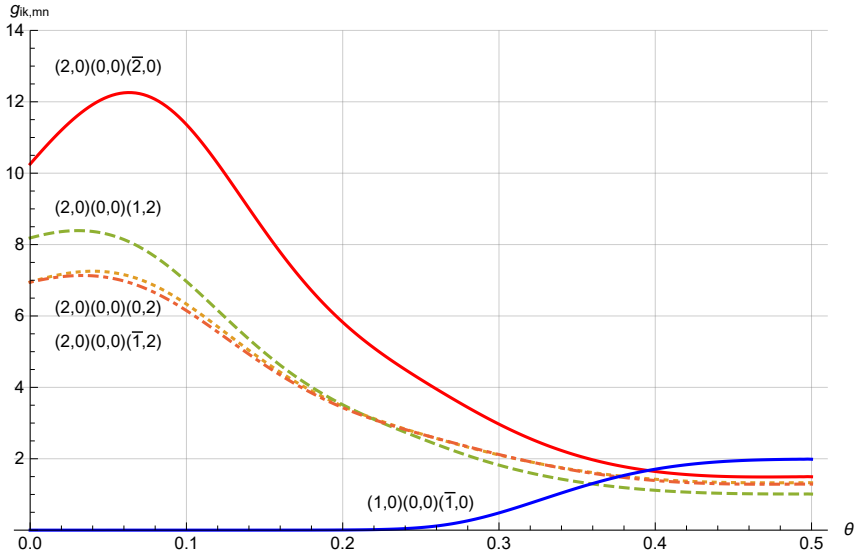

Fig. 4: Three-body distribution $g^{(3)}_{ik,mn}$, for 5 representative adatom trios $(i,k),(0,0),(m,n)$ at 400 K in dependence of coverage $\theta$, evaluated with Eq. (2.14), $g_{ik}(\theta)$ obeying Eq. (2.13). Red line: straight trios $(2,0)(0,0)(\bar{2},0)$, broken green line: 63° trios $(2,0)(0,0)(\bar{1},2)$, dotted line: 90° trios $(2,0)(0,0)(0,2)$, dot broken line: 116° trios $(2,0)(0,0)(\bar{1},2)$,, blue line: 45°straight n.n. trios $(1,0)(0,0)(\bar{1},0)$.

### 4.5. Temperature effects

Temperature has a strong influence on the pair distribution. In Eq. (2.13) the scaled interaction $u_{ik}=U_{ik}/k_BT$ is increased if the temperature is lowered.

As a result the peak value of $g_{20}$ near 0.1 ML coverage increases from 3.86 at 400 K to 5.45 at 350 K indicating a stronger tendency for ordering. An analog behavior shows the Fourier transform with maxima near 0.1 ML of 0.7 units at 350 K compared with 0.46 units at 400 K. At higher coverage e.g. at 0.4 ML there is a steep increase towards 0.7 units at 350 K compared with 0.25 units at 400 K.

At 500 K the $g_{20}$ peak is only 2.8 and the Fourier transform maximum is only 0.23 units. This indicates the onset of melting of the local order.

# 5. Discussion

In this section limitations and applicability and of the model are reviewed, the results of numerical calculations are discussed and aspects for further research are sketched.

### 5.1. Restrictions and limitations of the model

The results depend severely on the size of the 4th nearest neighbor interaction. A significant less attractive interaction would put p(2x2) formation in favor; its current size is related to the oscillating interaction derived in [10].

Due to surface reconstruction starting near half coverage the current approach is limited to the range below.

The pair distribution analysis makes use of Kirkwoods superposition approximation Eq. (2.2). The third step in the (infinite) distribution function hierarchy ignores the 4-party and higher distributions. So its reach clearly is limited to lower coverages; the absent of higher distributions also does not cover order fluctuations and estimates on their coherence length. The adatom-vacancy symmetry and the subsequent interpolation equation (2.13) extends its reach. The interpolation scheme, however, is not unique, others would lead to different results, but the qualitative picture would hold.

The interaction model is based on the theory of elasticity in the substrate and on lateral isotropic stress adatoms apply to the surface. Adatoms sitting on fourfold adatom sites creating isotropic stress are assumed. The electronic binding energy of adatom pairs and dipole-dipole interactions are ignored in the present model.

A further key assumption is an ideal flat surface, i.e. the absence of steps which are known for their significant attractive or repulsive interaction with adatoms.

The present model parameters rely on the DFT calculations in [13]. The adatom configuration base and their DFT energies used in this paper are restricted to the 5 configurations published there and thus are not at all exhaustive, but allow the conclusion of the actual model parameters [10]. A broader DFT base could change the model parameters and in consequence the results of both, the MC calculations, and the analytical pair distributions.

5.2. Consistency checks

Several consistency checks have been performed for the pair distribution algorithm:

- The interaction Eq. (3.3) was replaced by an isotropic $s^{-3}$ interaction. The resulting pair distribution was isotropic and showed a growing peak towards small distances *s* when the coverage was increased [22].

- The summation range in Eq. (2.13) was varied to check convergence.

- The principle value at $s_{14} \to 0$ in Eq. (2.13) was checked to vanish.

- After each coverage step the right hand side and the left hand side of Eq. (2.13) was compared. The differences were small, any significant differences would have caused instability of the algorithm.

- The temperature was varied, 350 K turned out as limit for the algorithms stability - much larger computing resources than available may shift the limit.

- The interpolation scheme in Eq. (2.13) was modified using logistic functions. Trigonometric functions were kept to avoid further parameters. The value *n=3* in the $\text{Cos}^n$ term in Eq. (2.13) turned out to allow consistency with the MC simulations. A smaller value of *n* would increase the maximum of $g_{ik}$ and shift it to higher coverage.

5.3. The many-body challenge

The current restriction to adatom pair interactions leaves out the question of many-body interactions in the region beyond 0.3 ML coverage where nearest neighbors may create many-body interactions [10] though marginal in size.

Eq. (2.3) cuts the distribution hierarchy at 3. An extension to properly handle higher distributions is seen as an interesting challenge to Statistical Physics. It would allow to better describe the coverage range beyond 0.2 ML.

5.4. Calculation results

Utilization of the 2-dimensional version of the BGY equation [12] for calculating the adatom pair distribution provides insight complementary to the Monte Carlo simulation. Both Fourier analysis and three-body distribution of the analytical pair distribution show the near range order while the Monte Carlo results give examples how configurations look like.

Neither p(2x2) nor c(2x2) grids are visible in the present model at 400 K. Therefore the existence of different phases and consequently a critical temperature is put in question for the O-Pd(100) system. Figs. 1., 2., 3. and 4. are interpreted as follows: At low coverage 3rd n.n. and 4th n.n. adatoms nucleate at $(2,0)(0,0)(\overline{2},0)$ trios or short chains. With increasing coverage mixed 3rd n.n./4th n.n. clusters grow and form at about 0.25 ML a coherent glassy structure. In the coverage range above vacancies are filled with 2nd n.n. and n.n. adatoms. The respective LEED spots of Fig.3. are broad but clearly differ from an amorphous ring.

The lack of long range order predicted by the current elastic model corresponds with an Ising model including long range interactions [23].

At elevated temperatures the local order has the trend for melting.

The assumption of long range elastic interactions has proved to be an adequate example.

Even if the 4th nearest neighbor attraction would be small the analytical approach would be useful.

### 5.5. Open questions and further aspects

The rising pair distribution values for nearest neighbors close to half coverage indicates the limits and shortcomings of the interpolation scheme.

A major challenge seems to be an extension of the statistical physics base towards 4-body and higher distributions. This would allow to cover the higher coverage range better.

A theoretical (DFT) model to determine the magnitude of the stress parameter $P$ could determine the elastic adatom interaction directly.

Extension of first principles calculations beyond the short range would prove elastic field effects and hopefully would provide their solidification.

Substrate materials other than Pd with different elastic constants and different types of adatoms may be interesting in comparing theory with a broader range of experimental findings. Especially the role of 4th n.n. attraction (forming 63° trios) seems interesting to be analyzed.

With appropriate computing resources the MC simulations could be improved.

Additional LEED data would enable a better comparison between measurements and theory. Especially the details of the LEED patterns (height, width) would allow conclusions on the spatial range of e.g. p(2x2) fluctuations or domains.

The influence of glassy surface structures on their catalytic activity seems to be an interesting topic.

# 6. Summary

The pair distribution of O-Pd(100) was evaluated using a BGY-type statistical method and a long range analytical interaction model based on first principles calculations. The particle-vacancy symmetry of pair interactions was utilized to adapt the BGY-type integral equation. Power series expansion allows the numeric solution of the equation up to half coverage. Pair distributions of some nearest neighbors are shown in dependence of coverage for 400 K. Together with the associated three-body distributions they provide a glance on the adatom short-range order in thermal equilibrium. Backed by Monte Carlo simulations the model predicts adatom formation of 3rd and 4th nearest neighbors due to a significant 4th near neighbor attraction. Therefore long range ordered structures like p(2x2) or c(2x2) are not found. Comparison of the Fourier transformed correlation with LEED spot measurements shows qualitative agreement but differences in their interpretation. While LEED data were previously interpreted with ordered phases, the present analysis proposes glassy structures with small ordered fluctuations. Limitations of the model are analyzed and open questions are addressed.

## Acknowledgment

This work is dedicated to my grandchildren. Continuous support of my family is gratefully acknowledged. Thanks to the unknown referee of v1 for the suggestion of adding Monte Carlo simulation. Thanks to the unknown referee of v2 for fruitful comments and for introducing refs. [11] and [17].